\documentclass[10pt,twoside]{scrartcl}

\usepackage[utf8]{inputenc}
\usepackage[T1]{fontenc}
\usepackage{amsmath,bbm}
\usepackage{dsfont}
\usepackage{algorithm}
\usepackage{algpseudocode}
\usepackage{telprint}
\usepackage{siunitx}
\usepackage{booktabs}
\usepackage{multirow}
\usepackage[paperwidth=209.9mm,paperheight=272mm,left=1.6cm,top=1.72cm,right=1.6cm,bottom=2cm,headheight=0cm]{geometry}
\usepackage{titling}
\usepackage{multicol}
\usepackage{graphicx}
\graphicspath{{./figs/}}
\usepackage{mathptmx}
\usepackage[compact]{titlesec}
\usepackage[format=plain,font=it,labelfont=it]{caption}
\usepackage{eso-pic}

\AddToShipoutPicture*{\footnotesize\sffamily\raisebox{1cm}{\hspace{1.65cm}\fbox{\parbox{\textwidth}{\copyright~2016 IEEE. Personal use of this material is permitted. Permission from IEEE must be obtained for all other uses, in any current or future media, including reprinting/republishing this material for advertising or promotional purposes, creating new collective works, for resale or redistribution to servers or lists, or reuse of any copyrighted component of this work in other works.}}}}
\renewenvironment{figure}{\par\medskip\noindent\minipage{\linewidth}}{\endminipage\par\medskip}

\DeclareMathAlphabet{\pazocal}{OMS}{zplm}{m}{n}

\newcommand{\X}{\mathbf{X}}

\newcommand{\M}{\mathbf{M}}
\newcommand{\G}{\mathbf{G}}
\newcommand{\y}{\mathbf{y}}
\newcommand{\pot}{\boldsymbol{\Phi}}
\newcommand{\temp}{\mathbf{T}}
\newcommand{\T}{\mathbf{T}}

\newcommand{\F}{\mathrm{F}}

\newcommand*{\rv}{\vec{r}}
\let\div\relax
\DeclareMathOperator{\div}{\nabla \cdot}
\DeclareMathOperator{\grad}{\nabla}

\newcommand{\bw}{\mathrm{bw}}

\newcommand{\sur}{\mathrm{sur}}
\newcommand{\hyb}{\mathrm{hyb}}

\newcommand{\Msigma}{\mathbf{M}_{\sigma}}
\newcommand{\Mrhoc}{\mathbf{M}_{\rho c}}
\newcommand{\Mlambda}{\mathbf{M}_{\lambda}}
\newcommand{\transpose}{^{\top}} 
\newcommand*{\fitDivd}{\widetilde{\mathbf{S}}}
\newcommand*{\nullVec}{\mathbf{0}}
\newcommand*{\Q}{\mathbf{Q}}
\newcommand*{\figref}[1]{Fig.~\ref{#1}}
\newcommand*{\thresh}{\delta}

\makeatletter
\renewcommand{\maketitle}{\bgroup
\begin{flushleft}
      \setlength{\baselineskip}{0pt}
      \textbf{\fontsize{14pt}{16pt}\selectfont\@title}\\
      \vspace{5pt}
      \@author
    \end{flushleft}\egroup
}
\makeatother

\title{Determination of Bond Wire Failure Probabilities in Microelectronic Packages}
\preauthor{\begin{flushleft}}
\author{Thorben~Casper\textsuperscript{1,2,*}, Ulrich~Römer\textsuperscript{1,2}, Sebastian~Schöps\textsuperscript{1,2}}
\postauthor{\end{flushleft}}

\titleformat{\section}{\fontsize{11pt}{13pt}\selectfont\bfseries}{\thetitle\hspace{1.45em}}{0em}{}
\titleformat{\subsection}{\bfseries}{\thetitle\hspace{0.87em}}{0em}{}

\begin{document}
\maketitle
{\setlength\parskip{0pt}
\vspace{1pt}
\textsuperscript{1} Graduate School of Computational Engineering, Technische Universität Darmstadt, 64293 Darmstadt, Germany\\[-0.5pt]
\textsuperscript{2} Institut für Theorie Elektromagnetischer Felder, Technische Universität Darmstadt, 64289 Darmstadt, Germany\\
\phantom{dummy}\\[-1pt]
$\textrm{*}$ Corresponding Author: casper@gsc.tu-darmstadt.de, +49\,6151\,16\,24392\\

\vspace{13pt}
\textit{\fontsize{11pt}{1.2}\selectfont Abstract}
\vspace{5pt}
}

\textit{\fontsize{10pt}{0pt}\selectfont This work deals with the computation of industry-relevant bond wire failure probabilities in microelectronic packages. Under operating conditions, a package is subject to Joule heating that can lead to electrothermally induced failures. Manufacturing tolerances result, e.g., in uncertain bond wire geometries that often induce very small failure probabilities requiring a high number of Monte Carlo (MC) samples to be computed. Therefore, a hybrid MC sampling scheme that combines the use of an expensive computer model with a cheap surrogate is used. The fraction of surrogate evaluations is maximized using an iterative procedure, yielding accurate results at reduced cost. Moreover, the scheme is non-intrusive, i.e., existing code can be reused. The algorithm is used to compute the failure probability for an example package and the computational savings are assessed by performing a surrogate efficiency study.}

\vspace{1em}

\begin{multicols}{2}

\section{Introduction}
In nowadays' micro- and nanoelectronic applications, constant downscaling leads to increasing power densities.
Arising thermal problems can trigger the degeneration of materials, performance restrictions or even system failure.
Typically, thermal designers use guidelines to avoid thermal problems in their designs. This may lead to inaccurate modeling and overdesign. Additionally, uncertainties in material and geometrical properties stemming from tolerances in the manufacturing process may result in unexpected behavior.

For a more accurate prediction of the manufactured functionality, numerical simulations are becoming increasingly popular.
Using relevant (uncertain) parameters as inputs, uncertainty quantification techniques help to understand the influence of manufacturing tolerances on the devices' performance.
The evaluation of failure probabilities becomes possible, yet the computation of very small failure probabilities is numerically challenging. These small probabilities inevitably occur in the context of a six sigma design goal.

\begin{figure}
    \centering
    \includegraphics[width=0.55\columnwidth]{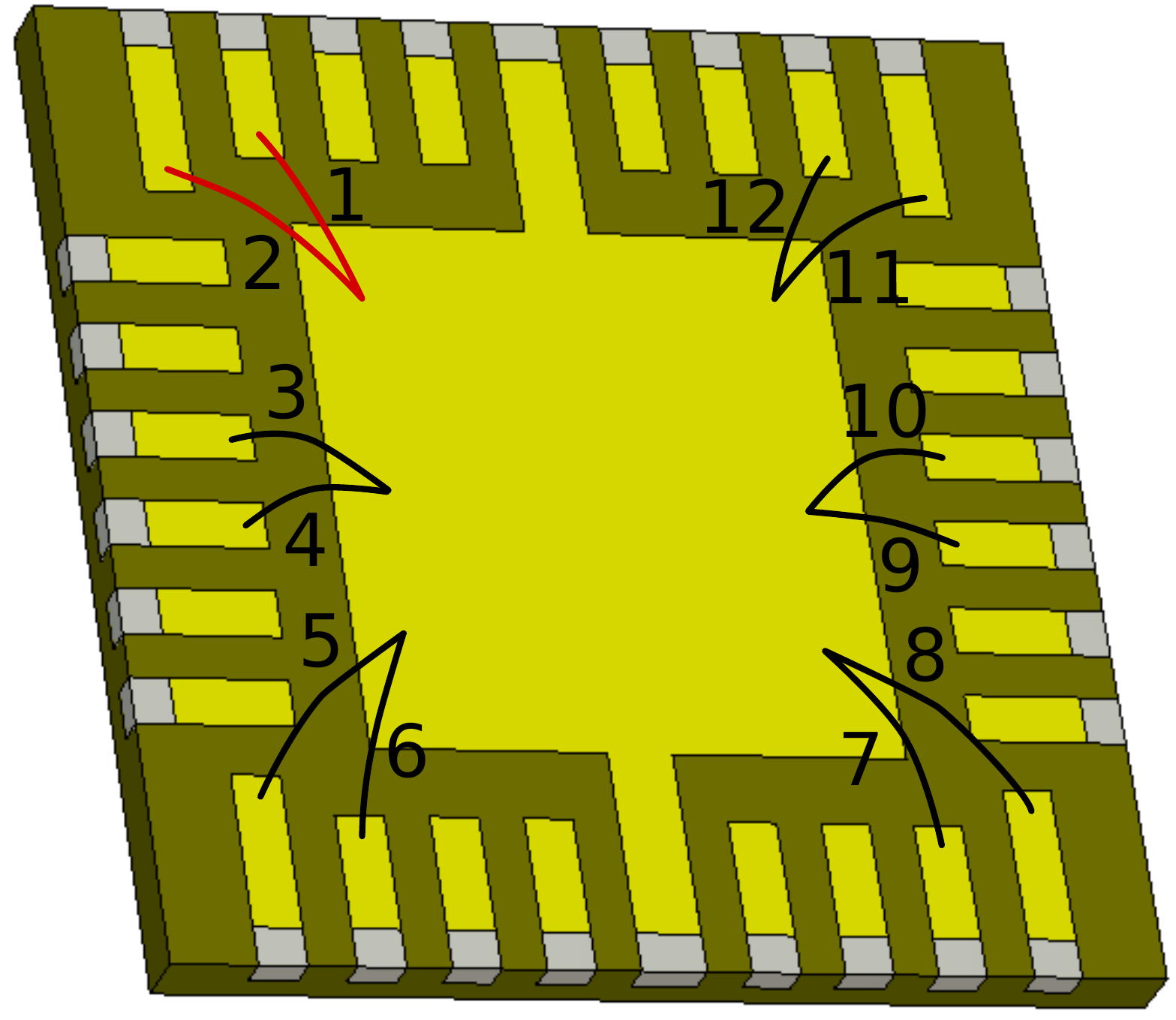}
    \captionof{figure}{Microelectronic chip package with bond wires. For the study presented here, the geometry of bond wires \num{1} and \num{2} (red) is modeled to be subject to uncertainty while the parameters of the other wires are assumed to be known exactly.}
    \label{fig:chip}
\end{figure}

One possible approach to compute failure probabilities was first presented by Li and Xiu in~\cite{Li_2010aa} and uses combined Monte Carlo sampling of the original computational model with a polynomial approximation.
The idea is to use the computationally cheaper but less accurate polynomial model as long as a sample far away from the failure region is considered.
Once a sample falls within the vicinity of the failure region, the original expensive model is used to ensure accurate results.
The threshold, deciding which model needs to be evaluated, is determined iteratively as outlined in~\cite{Li_2010aa}.
A possible extension to additionally compute rare failure probabilities, i.e., below $10^{-5}$, can be realized by, e.g., the usage of importance sampling as presented in~\cite{Li_2011ab}. To estimate the threshold a posteriori by using an adjoint error estimator, see~\cite{Roemer_2017aa}.

In this paper, the method presented in~\cite{Li_2010aa} is applied to compute bond wire failure probabilities in a microelectronic chip package as shown in \figref{fig:chip}. The paper is organized as follows. First, Section~\ref{sec:electrothermal_problem} introduces the underlying electrothermal problem in the continuous and the discrete setting including the bond wire contribution. Then, different approaches to compute the failure probability are presented in Section~\ref{sec:failure_probability}, before the numerical results are given in Section~\ref{sec:simulation_results}. Finally, Section~\ref{sec:conclusion} concludes the paper.

\medskip
\section{Electrothermal Problem}
\label{sec:electrothermal_problem}

One of the main reasons why bond wires are subject to failure is because of the Joule heating effects that stem from applied currents. 
For the evaluation of failure probabilities, it is thus required to analyze the coupled electrothermal system. 
\medskip
\subsection{Continuous Setting}

Disregarding transient effects in the electrical problem, we consider the coupling of the electrokinetic problem with the transient heat equation. With the computational domain $\Omega$, $\rv\in\Omega$ and ${t\in I=\left(0,t_\text{end}\right]}$ being the coordinates in space and time, respectively, the continuous setting is given by

\begin{subequations}
    \begin{align}
        - \div \left(\sigma(\rv,T)\grad \varphi(\rv,t) \right) &= 0,  \\
        \rho(\rv) c(\rv) \dot{T}(\rv,t) - \div \left(\lambda(\rv,T)\grad T(\rv,t) \right) &= Q(\varphi,T),\label{eq:contHeatEq}
    \end{align}\label{eq:electrothermal}
\end{subequations}
with appropiate initial and boundary conditions. Note that the time dependence of the electric potential $\varphi$ is induced by the coupling with the transient heat equation.
In~\eqref{eq:contHeatEq}, the Joule loss coupling term ${Q_\text{el}=\sigma\left|\nabla\varphi \right|^2}$ is incorporated as a contribution to the heat power density $Q$.
The material parameters are given by the electrical conductivity $\sigma$, the volumetric heat capacity $\rho c$ and the thermal conductivity $\lambda$. While we neglect the temperature dependence of $\rho$ and $c$, we model $\sigma$ and $\lambda$ to be functions of temperature. The contribution of the bond wires (cf.~\figref{fig:chip}) is omitted for now and will be included in the discrete setting in Section~\ref{sec:bw_contribution}.

\medskip
\subsection{Discrete Setting}

For the solution of~\eqref{eq:electrothermal}, a numerical scheme is required. Here, we choose the Finite Integration Technique (FIT) \cite{Weiland_1996aa,Clemens_2001ac} on a pair of hexahedral meshes to obtain the semi-discrete system
\begin{subequations}
    \begin{align*}
        -\fitDivd\Msigma(\T)\G \pot &= \nullVec, \\
        \Mrhoc\dot{\T} - \fitDivd\Mlambda(\T)\G \T &= \Q(\pot,\T).
    \end{align*}
\end{subequations}
The time-dependent degrees of freedom are the potential vector $\pot$ and the temperature vector $\T$. In analogy to the continuous problem, the Joule heating contribution $\Q_\text{el}$ adds to the vector of source heat powers $\Q$.
The materials are given by the electric conductance matrix $\M_{\sigma}$ and the thermal capacitance and conductance matrices $\M_{\varrho c}$ and $\M_{\lambda}$, respectively.
The dual face to volume incidence matrix $\fitDivd$ and the primary node to edge incidence matrix $\G = -\fitDivd\transpose$ are the discrete analogons to the continuous divergence and gradient operator, respectively. 
Subsequent time discretization is done using the implicit Euler method together with a fractional step splitting for the algebraic equation.

\medskip
\subsection{Bond Wire Contribution}
\label{sec:bw_contribution}

Since the extent of bond wires is very small compared to the remaining feature sizes in a chip package, the wires are not resolved in the mesh but rather modeled by a lumped element approach. To include the wire contribution in the discrete setting, a stamping approach as outlined in \cite{Casper_2016aa} is applied. Then, the discrete system including $N^\bw$ bond wires reads
\begin{subequations}
    \begin{align*}
        \fitDivd\Msigma(\T)\fitDivd\transpose\pot+\sum_{j=1}^{N^\mathrm{bw}} \mathbf{P}_{j}G^{\text{bw},j}_{\text{el}}(T^{\text{bw},j})\mathbf{P}_{j}\transpose\pot &= \mathbf{0}, \\
        \Mrhoc \dot{\T} + \fitDivd\Mlambda(\T)\fitDivd\transpose \T + \sum_{j=1}^{N^\mathrm{bw}} \mathbf{P}_{j}G^{\text{bw},j}_{\text{th}}(T^{\text{bw},j})\mathbf{P}_{j}\transpose{\T} &= \hat{\mathbf{Q}}(\pot,\T),
    \end{align*}
\end{subequations}
where $\mathbf{P}_{j}$ is the incidence vector between the bond wire contacts and the dual volumes, containing entries ${0,-1}$~and~$1$. The temperature $T^{\text{bw},j}=\X_j\transpose\T$ of a bond wire, with ${\X_j = \frac{1}{2}|\mathbf{P}_j|}$, where $|\cdot|$ refers to the vector of absolute values, is defined as the average value of the temperature at its end points. The source term $\hat{\mathbf{Q}}$ comprises Joule heating of both the distributed part $\Q_\text{el}$ and bond wire part as
\begin{equation*}
    \hat{\mathbf{Q}}(\pot,\T) = \mathbf{Q}_\text{el}(\pot,\T) + \sum_{j=1}^{N^\mathrm{bw}} \X_j G^{\text{bw},j}_{\text{el}}(T^{\text{bw},j})( \pot\transpose\mathbf{P}_{j})^2,
\end{equation*}
with the electrical and thermal conductance of bond wire~$j$ given by $G^{\text{bw},j}_{\text{el}}$ and $G^{\text{bw},j}_{\text{th}}$, respectively. Here, $\X_j$ distributes the heat generated in the bond wire to the dual volumes to which the bond wire is connected. In this paper, we model each bond wire with a length $l_j$ and a uniform cross section $A_j$. Therefore, the conductance of wire $j$ reads ${G^{\text{bw},j}_{\text{\{el,th\}}} = \{\sigma,\lambda\}A_j/l_j}$.

\medskip
\section{Failure Probability}
\label{sec:failure_probability}

Failures occurring in technological applications due to manufacturing tolerances (resulting in, e.g., uncertain geometries) are of stochastical nature. 
In the framework of a microelectronic chip package with bond wires connecting the chip with its package, we define a failure as the fusing or breaking of a bond wire. 
We thereby assume that failures originate predominantly from the bond wire and thus neglect any other possible sources for failure. The considered uncertain bond wire geometries shall be modeled by the realization $\y$ of a random variable in the observation space $\Gamma\subset\mathbbm{R}^N$. Then, the electric and thermal conductance $G_{\{\text{el},\text{th}\}}^{\bw,j}$ depends on $\y$ and so do $\pot$ and $\temp$, implicitly, through the electrothermal problem.

The evaluation of the associated failure probabilities is divided into the following steps. We give a mathematical discussion of the failure of a single bond wire and present methods to compute the associated failure probability. Afterwards, system failure probability is defined based on the individual bond wire failure probabilities.

\medskip
\subsection{Bond Wire Failure}
A single bond wire failure induced by stress, electromigration or other reasons shall be modeled by a critical temperature $T_\text{cr}$ that we assume to be equal for all here considered wires. Then, a failure occurs when the maximum temperature of the wire exceeds the critical temperature $T_\text{cr}$ at any instant in time, given by 
\begin{equation}
    \max_{t\in I} T^{\bw}\left(t,\y\right) > T_\text{cr},
    \label{eq:bwFailure}
\end{equation}
where the wire's temperature $T^{\bw}$ now depends on $\y$.
Following~\cite{Li_2010aa}, the failure shall be modeled with the help of a performance function $g: \mathbbm{R}^N \rightarrow \mathbbm{R}$ that describes a failure of the wire when $g<0$. A possible choice for this function is
\begin{equation*}
    g(\y) = -\max_{t\in I} T^{\bw}\left(t,\y\right) + T_\text{cr}.
\end{equation*}

\begin{figure}
    \centering
    \includegraphics[width=.48\textwidth]{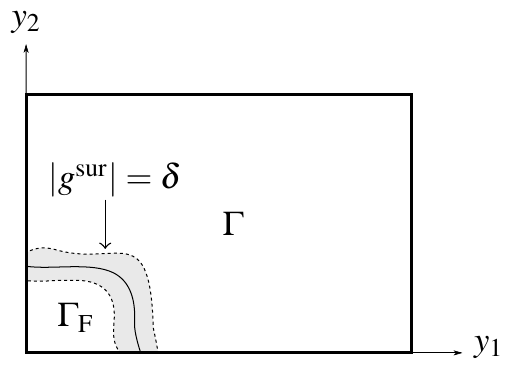}
    \captionof{figure}{Separation of the domain $\Gamma$ into a region where the full model is evaluated (gray) and where the surrogate model is evaluated (white). Full model evaluations within the gray area are triggered by the threshold through $|g^\sur|=\thresh$.}
    \label{fig:gamma_region}
\end{figure}

Those points of $\Gamma$ resulting in a negative performance function constitute the failure region, denoted as $\Gamma_\F$. For the case of two uncertain input variables $y_1$ and $y_2$, \figref{fig:gamma_region} depicts schematically the rectangular observation space $\Gamma$ and its subspace $\Gamma_\F$, separated by the solid line. With the definition of the characteristic function
\[
    \mathbbm{1}_{\Gamma_\F}(\y) = 
    \left \{
        \begin{aligned}
            &1, && \y \in \Gamma_\F, \\
            &0, && \y \notin \Gamma_\F, \\
        \end{aligned}
        \right .
    \]
    the failure probability is given by
    \begin{equation*}
        P_\F = \int_\Gamma \mathbbm{1}_{\Gamma_\F}\left(\y\right)\varrho\left(\y\right)\,\text{d}\y.
    \end{equation*}

\subsection{Sampling Scheme}

    Commonly, failure probabilities are approximated using Monte Carlo sampling yielding
    \begin{equation*}
        P_\F \approx \frac{1}{M} \sum_{i=1}^M \mathbbm{1}_{\Gamma_\F}(\y^i),
    \end{equation*}
    with $M$ being the number of Monte Carlo samples and $\y^i$ a random sample of $\y$ drawn according to the probability distribution~$\varrho$. The complexity of this approach highly depends on the cost to evaluate $\mathbbm{1}_{\Gamma_\F}(\y^i)$. 

\medskip
\subsection{Surrogate Model}

    The problem for the sampling scheme described in the previous section is that for each sample, the evaluation of~$g$ requires the solution of a system of PDEs, i.e., system~\eqref{eq:electrothermal}. 
    Therefore, especially for small failure probabilities, a large number of samples $M$ is required and an efficient surrogate model is necessary. 

    A class of polynomial surrogate models, referred to as generalized polynomial chaos, was proposed in \cite{Xiu_2002aa}. These global polynomials are at the core of spectral stochastic methods, such as the stochastic Galerkin or collocation method. Here, the non-intrusive collocation procedure presented in \cite{Babuska_2007aa} is adopted, as it is readily applicable to the present nonlinear, transient and coupled problem. 
    In the simplest case of tensor grid collocation, with collocation points $\hat{\y}^m$, ${m=1,\ldots,(p+1)^N}$, with polynomial degree $p$, the model is approximated as
    \begin{equation*}
        g(\y) \approx g^\sur(\y)  = \sum_{m=1}^{(p+1)^N} g(\hat{\y}^m) L_m(\y),
    \end{equation*}
    where $L_m$ are multivariate Lagrange polynomials. However, when a high number of uncertain input parameters is involved, more sophisticated methods such as sparse grids \cite{Nobile_2008aa} and low-rank tensor approximations \cite{Loukrezis_2016aa} are used.  

    In the following, the failure region evaluated with $g^\sur$ instead of $g$ is called $\Gamma_\F^\sur$ and the associated failure probability reads
    \begin{equation}
        P_\F^\sur = \int_\Gamma \mathbbm{1}_{\Gamma_\F^\sur}\left(\y\right)\varrho\left(\y\right)\,\text{d}\y.
        \label{eq:failureProbSurrogate}
    \end{equation}
    With this surrogate model, only polynomial evaluations are required, reducing the computational cost substantially. 

\subsection{Hybrid Scheme}
\label{sec:hybrid_scheme}

    While the evaluation of the presented surrogate model is more efficient than the evaluation of the full model, the computed failure probabilities can be inaccurate. It has been shown, that this shortcoming may persist even if a very accurate surrogate model is employed together with a large number of samples~\cite{Li_2010aa}. 
    As a remedy, a hybrid scheme, combining the accurate PDE model with the efficient surrogate model is used~\cite{Li_2010aa}.

    The idea is to evaluate the surrogate model only when the considered sample lies far away from the boundary of the failure region, i.e., $|g^\sur(\y^i)|$ larger than a certain threshold $\thresh$.
    Once $|g^\sur(\y^i)|$ becomes smaller than this threshold and thus close to the failure region, the original model is used to evaluate $\mathbbm{1}_{\Gamma_\F}$ instead of $\mathbbm{1}_{\Gamma_{\F}^\sur}$. 
    This idea is illustrated in \figref{fig:gamma_region} and reduces the problem to the one of finding an adequate value for~$\thresh$. 

    One possible approach to determine $\thresh$ is an iterative method~\cite{Li_2010aa} as outlined in the following. First, the surrogate model is evaluated for the full set of samples to obtain $\{g^\sur(\y^i)\}_{i=1}^{M}$. For each $g^\sur(\y^i)$, $\mathds{1}_{\Gamma_\F^\sur}$ can be evaluated with implicitely given $\Gamma_\F^\sur$ and hence the hybrid failure probability $P_\F^{\hyb,(1)} = P_\F^\sur$ is computable. Then, the full model is evaluated for the $\delta M < M$ samples that are the closest to the failure region. For these samples $\mathds{1}_{\Gamma_\F}^\sur$ is replaced with $\mathds{1}_{\Gamma_\F}$ in the failure probability estimate. If the associated change in the failure probability estimate exceeds a certain tolerance $\eta$, the failure probability $P_\F^{\hyb,(k+1)}$ is updated with the $\delta M$ full model evaluations. Then, the next $\delta M$ samples are chosen and the procedure is repeated until $|P_\F^{\hyb,(k+1)} - P_\F^{\hyb,(k)}| \leq \eta$. The here described procedure is given in Algorithm~\ref{alg:hybrid}, where it is understood that $M/\delta M$ is an integer.

    This algorithm ensures that the error in the computed failure probability decreases as the accuracy of the surrogate model increases, in contrast to sampling the surrogate model solely~\cite{Li_2010aa}. However, the result might depend on the choice of the stepsize $\delta M$, which has to be determined empirically. As an alternative to this iterative approach, the value of the threshold $\delta$ is estimated a posteriori using an adjoint approach in~\cite{Roemer_2017aa}, yielding full control on the accuracy of the failure probability estimate.

    Mathematically, the resulting hybrid failure probability is given by
    \begin{equation*}
        P_\F^\hyb = \int_\Gamma \mathbbm{1}_{\Gamma_\F^\hyb}(\y) \varrho\left(\y\right)\,\text{d} \y,
    \end{equation*}
    with
    \begin{equation*}
        \mathbbm{1}_{\Gamma_\F^\hyb} = \mathbbm{1}_{\{g^\sur<-\delta\}} + \mathbbm{1}_{\{\left|g^\sur\right|<\delta\}\cap\{g<0\}}.
    \end{equation*}
    Here, the compact notation of e.g.\ $\{g<0\}$ is short for $\{\y\ |\ g(\y) < 0\}$. 

\subsection{System Failure}
\label{sec:system_failure}

    In the previous sections, we introduced a method to compute the failure probability of bond wires efficiently. If a system of $N^\bw$ bond wires is considered, the system fails if any of these wires fails. Therefore, with~\eqref{eq:bwFailure}, the condition for system failure is given by
    \begin{equation*}
        \max_{j=1,\dots,N^\bw}\max_{t\in I} T^{\bw,j}\left(t,\y\right) > T_\text{cr}.
    \end{equation*}

    It has been observed \cite{Loukrezis_2016aa} that taking the maximum over all bond wires gives rise to a performance function that is not smooth and hence difficult to approximate with polynomials. For an efficient surrogate approximation, we thus calculate the failure probability $P_{\F,j}^{\text{hyb}}$ for every wire independently. This also requires the repetitive application of Algorithm~\ref{alg:hybrid} for every single wire, as the $\{\y^i\}_{i=1}^M$ are resorted in a different order for each wire. The event of failure of a single wire is not disjoint from the event of failure of another wire in general. 

Hence, we deduce from the basic axioms of probability theory, that the system failure probability can be estimated as
    \begin{equation}
        P_{\F,\text{s}}^{\text{hyb}} \leq \sum_{j=1}^{N^\bw}P_{\F,j}^{\text{hyb}}.
        \label{eq:system_failure}
    \end{equation}

\vspace{1ex}
\begin{algorithmic}[1]
    \Procedure{hybrid}{$g$, $g^\sur$, $M$,$\delta M$,$\y$,$\eta$}
    \State set $M^{(1)}= 0$\Comment{Initialization}
    \State evaluate $P_\F^{\hyb,(1)} = P_\F^\sur$ using \eqref{eq:failureProbSurrogate}
    \State sort $\{\y^i\}_{i=1}^M$ as $\{\y_\text{asc}^i\}_{i=1}^M$ s.t. $\{|g^\sur(\y_\text{asc}^i)|\}_{i=1}^M$ ascends
    \For{$k = 1\ \textbf{to}\ M/\delta M $}\Comment{Iteration}
    \State define $M_1=M^{(k)}+1$ 
    \State define $M_2=M^{(k)}+\delta M$
    \State evaluate $g(\{\y_\text{asc}^i\}_{i=M_1}^{M_2})$ yielding $\mathds{1}_{\Gamma_\F}(\{\y_\text{asc}^i\}_{i=M_1}^{M_2})$
    \State set $\Delta P_\F^{\hyb,(k)} = \frac{1}{M} \sum_{i=M_1}^{M_2}\left(-\mathds{1}_{\Gamma_\F^\sur}(\y_\text{asc}^i) +  \mathds{1}_{\Gamma_{\F}}(\y_\text{asc}^i) \right)$
    \State set $P_\F^{\hyb,(k+1)} = P_\F^{\hyb,(k)} + \Delta P_\F^{\hyb,(k)}$
    \If{$|P_\F^{\hyb,(k+1)} - P_\F^{\hyb,(k)}| \leq \eta$}
    \State \textbf{return} $P_\F^{\hyb,(k+1)}$
    \EndIf
    \State update $M^{(k+1)} = M^{(k)} + \delta M$
    \EndFor\label{euclidendwhile}
    \State \textbf{return} $P_\F^{\hyb,(k+1)}$
    \EndProcedure
\end{algorithmic}
\captionof{algorithm}{Iterative algorithm to compute the hybrid failure probability using an expensive computer model $g$ and a cheaper surrogate model $g^\text{sur}$. The tolerance $\eta$ determines when the algorithm terminates.}\label{alg:hybrid}

\section{Simulation Results}
\label{sec:simulation_results}
    In this section, we apply Algorithm~\ref{alg:hybrid} to the example of a microelectronic chip package including bond wires of uncertain geometry (see \figref{fig:chip}).
    The goal of a bond wire designer is to dimension the bond wires such that the failure probability is low while minimizing the overall cost. For simplicity, we assume that the lengths of the wires are predetermined by the package, leaving the designer with the wires' diameter as the design parameter. Neglecting aging effects, the uncertain quantities $\y$ are modeled to be the relative change of the bond wire lengths. 
    Therefore, we are assuming that the diameter of a wire is precisely controlled by the manufacturing process and a wire's geometry is only subject to an uncertain length $l_i = l_{i,0}/(1-y_i)$, with a deterministic length $l_{i,0}$~\cite{Casper_2016aa}. 

\medskip
\subsection{Numerical Setting}

The here used setting for the simulation of the microelectronic chip package has been presented in \cite{Casper_2016aa}. In this paper, for simplicity and as shown in \figref{fig:chip}, we choose bond wires $1$ and $2$ to be subject to an uncertain length. Therefore, $N=2$ and the setup of the surrogate model is simplified. We can then focus on the iterative hybrid sampling algorithm which is the main topic of this work. 
For the high-dimensional case $N=N^\bw=12$, we refer to~\cite{Roemer_2017aa} and~\cite{Loukrezis_2016aa}. Since there is not sufficient measurement data available to determine the probability density function, we choose $y_i$ to be uniformly distributed in the interval $[\mu - \sigma, \mu + \sigma]$ with $\mu=0.17$ and $\sigma=0.048$.

If not otherwise stated, a first order surrogate ($p=1$) with $M=\num{1e5}$ samples was used. The tolerance to determine the termination of the hybrid algorithm is chosen to $\eta=\num{1e-10}$.
For the bond wires' geometry, we assume that a designer has chosen the diameter to ${d=\SI{12.07}{\micro\metre}}$ and analyze this setting with the here presented method.
The given simulation settings are also summarized in Table~\ref{tab:simulation_setting}.

\captionof{table}{Simulation settings.}\label{tab:simulation_setting}
\vspace{-3ex}
\begin{center}
\begin{tabular}{|clr|}\hline
    Symbol & \multicolumn{1}{c}{Description} & \multicolumn{1}{c|}{Value}\\ \hline
    $N^\bw$ & No.\ of wires in the model & \num{12} \\
    $d$ & Wires' diameter & \SI{12.07}{\micro\metre}\\ 
    $h$ & Heat transfer coefficient & \SI{25}{W/m^2/K}\\
    $N_t$ & No.\ of time steps & \num{51} \\
    $t_\text{end}$ & End time & $\SI{50}{s}$\\
    $V_{\text{bw}}$ & Bond wire voltage & \SI{40}{mV}\\ 
    $T_\infty$ & Ambient temperature & \SI{300}{K}\\
    $T_\text{cr}$ & Critical temperature & \SI{523}{K}\\
    $p$ & Polynomial degree of surrogate & \num{1} \\
    $N$ & No.\ of uncertain wires & \num{2} \\
    $M$ & No.\ of samples & \num{1e5} \\
    $\eta$ & Tolerance for hybrid algorithm & \num{1e-10} \\ \hline
\end{tabular}
\end{center}
\vspace{1ex}

\subsection{Heating of the Chip}

Due to the constant applied voltage and the convective thermal boundary conditions, the chip heats up until a stationary state is reached. \figref{fig:TbwireMaxVStime} depicts the expected value of the temperature (blue) of bond wire $9$ (cf. \figref{fig:chip}) based on the hybrid approach with $\delta M = \num{10}$ as presented in Section~\ref{sec:hybrid_scheme}. 

The (red) horizontal line shows the critical temperature $T_\text{cr}$ as a reference for failure. Furthermore, error bars showing the $6\sigma$-deviation are plotted. At $t=t_\text{end}=\SI{50}{s}$, the standard deviation is given by $\sigma_\text{MC} = \SI{1.10}{K}$.

\medskip
\subsection{Hybrid Failure Probability}

\begin{figure}
  \centering
  \includegraphics[width=.8\textwidth]{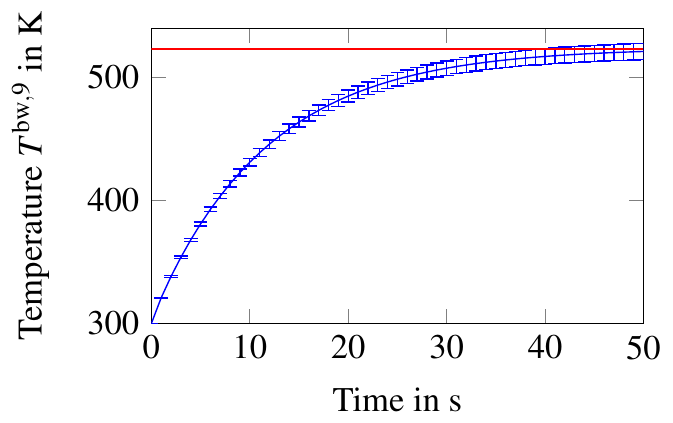}
  \captionof{figure}{Temperature of bond wire $9$ over time with $6\sigma$ variation (blue) evaluated using the hybrid approach with $\delta M = \num{10}$. The horizontal (red) line shows the critical temperature $T_\text{cr}$ as a reference for failure.}
  \label{fig:TbwireMaxVStime}
\end{figure}

From \figref{fig:TbwireMaxVStime}, we see that a small but nonzero failure probability is expected since the $6\sigma$ deviation crosses the reference line for a wire failure. Note that in proper six sigma design, the upper limit of the six sigma interval would be expected to be entirely below the red line. However, in the present setting, the increased failure probability simplifies numerical investigations of the hybrid iterative algorithm.

We recall that a quantification of the failure probability based on the surrogate model solely may be inaccurate. Hence, Algorithm~\ref{alg:hybrid} is applied here. We compute the failure probability of each wire separately using different values for $\delta M$. Since the geometry of wires \num{9} and \num{10} ensures a higher conductance than all other wires, the highest temperatures are observed for these two wires. The here presented example is chosen such that the failure probability of these wires is very small. As the temperature of all other wires is lower, the failure probability is lower as well. However, the here chosen number of samples does not resolve these even smaller failure probabilities and are therefore computed to be zero.
 
\begin{figure}
    \centering
    \includegraphics[width=.95\textwidth]{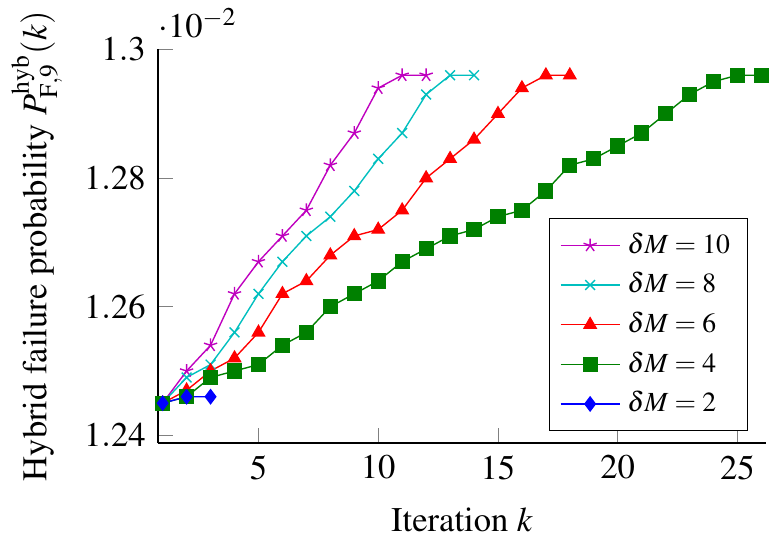}
    \captionof{figure}{Hybrid failure probability $P_\F^{\mathrm{hyb}}$ over iteration $k$ for bond wire $9$ and different choices of $\delta M$.}
    \label{fig:PfhybVSiterationWire9}
\end{figure}

\begin{figure}
    \centering
    \includegraphics[width=.95\textwidth]{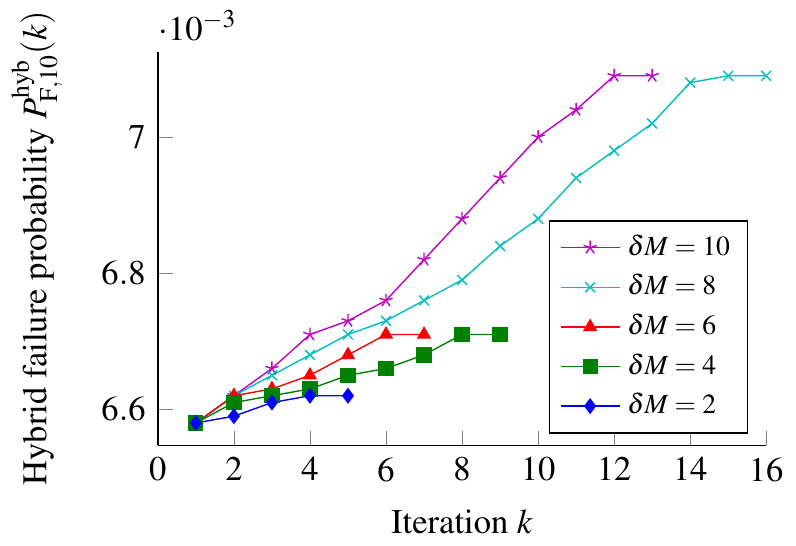}
    \captionof{figure}{Hybrid failure probability $P_\F^{\mathrm{hyb}}$ over iteration $k$ for bond wire $10$ and different choices of $\delta M$.}
    \label{fig:PfhybVSiterationWire10}
\end{figure}

The resulting evolutions of $P_\F^\hyb$ of wire $9$ and $10$ (cf. \figref{fig:chip}) are shown in \figref{fig:PfhybVSiterationWire9} and~\ref{fig:PfhybVSiterationWire10}, respectively. The converged failure probability for wire \num{9} is $P_{\F,9}^{\hyb}\approx\num{0.013}$ and the one for wire \num{10} is $P_{\F,10}^{\hyb}\approx\num{0.0071}$. It can be observed that the calculated failure probability changes only slightly during the execution of the algorithm. The reason for the small variation is that a change in the failure probability only occurs if a sample falls within the region where the surrogate model does not compute the failure of the wire correctly. Since this region is apparently very small compared to the failure region $\Gamma_\F$, there is only a very small but nonzero change observed. 
Additionally, we note that the number of required iterations until convergence depends highly on the chosen value for $\delta M$. Moreover, the algorithm might not converge at all if $\delta M$ is chosen too small. This is e.g.\ observed for $\delta M = 2$ in the case of wire \num{9} and even for $\delta M = \{2,4,6\}$ for wire \num{10} (cf. \figref{fig:PfhybVSiterationWire9} and~\ref{fig:PfhybVSiterationWire10}). Hence, to ensure the accuracy of the final probability estimate, repetitive runs of Algorithm~\ref{alg:hybrid} with different choices of $\delta M$ are recommended.

Apart from the failure probability of a single wire, the system failure probability was defined in Section~\ref{sec:system_failure}. With~\eqref{eq:system_failure}, it can be estimated for the here considered values of $\delta M$. The results range between \num{0.0191} and \num{0.0200}. However, we recall that the results for $\delta M = \{2,4,6\}$ have not converged as observed from \figref{fig:PfhybVSiterationWire9} and~\ref{fig:PfhybVSiterationWire10}. As already mentioned, the failure probability of all other wires was computed to zero since none of the Monte Carlo samples falls neither in the failure region nor in the threshold region defined by $\delta$. The algorithm for these wires therefore converges directly after the first iteration. The results for the system failure probability are summarized in Table~\ref{tab:hybridSystemProb}.

\vspace{1ex}

\vspace{-3ex}
{
\begin{center}
\captionof{table}{System failure probability $P_{\F,\text{s}}^\hyb$ for different $\delta M$.}\label{tab:hybridSystemProb}
\vspace{0.5ex}
\begin{tabular}{|c|ccccc|}\hline
    $\delta M$ & \num{2} & \num{4} & \num{6} & \num{8} & \num{10} \\ \hline
    $P_{\F,\text{s}}^\hyb$ & \num{0.0191} & \num{0.0197} & \num{0.0197} & \num{0.0200} & \num{0.0200} \\ \hline
\end{tabular}
\end{center}
}

\medskip
\subsection{Efficiency of Surrogate Model}

The surrogate is constructed using a tensor grid with ${(p+1)^N}$ collocation points for which the full model needs to be computed. Therefore, the effort to construct the surrogate model depends on the number of uncertain parameters and on its polynomial degree. 
Furthermore, the number of required iterations to execute Algorithm~\ref{alg:hybrid} depends on the choice of $\delta M$ as it is visible from \figref{fig:PfhybVSiterationWire9} and~\ref{fig:PfhybVSiterationWire10}. In Table~\ref{tab:surrogateEfficiency}, the total number of full model calls to calculate the hybrid failure probability of wires \num{9} and \num{10} is given as a function of the polynomial degree $p$ and the iteration parameter $\delta M$. Comparing surrogate degree \num{1} and \num{2} under the condition that convergence was observed for the chosen $\delta M$, it is seen that the accuracy of the higher order surrogate leads to less iterations and therefore less full model calls. Furthermore, for degree \num{3}, it is observed that the algorithm terminates directly after only one iteration.

The cost of using a surrogate with $p=3$ is comparable to the case $p=2$. However, the hybrid algorithm with surrogates of degree $p>3$ require a more expensive setup without giving more accurate results. Moreover, it shall also be noted that all presented combinations lead to a much cheaper computation of the failure probability than using a pure Monte Carlo sampling with $M=\num{1e5}$ full model evaluations.

\medskip
\section{Conclusions}
\label{sec:conclusion}
The hybrid scheme first presented in~\cite{Li_2010aa} has been applied to the evaluation of bond wire failure probabilities for the example of a microelectronic chip package as it has been presented in~\cite{Casper_2016aa}.
For a particular wire and a Monte Carlo sampling of a first order surrogate model ($p=1$) with \num{1e5} samples, the temperature as the result of the hybrid algorithm has been presented as a function of time. Since the six sigma deviation of this computed temperature exceeds the critical 
temperature $T_\text{cr}$, a non-zero failure probability was expected.
{
\begin{center}
\captionof{table}{Number of full model calls to compute the hybrid failure probabilities for wires $9$ and $10$ in dependence of the surrogate polynomial level $p$ and the iteration parameter $\delta M$. The number of model calls to set up the surrogate model is included in the data.}\label{tab:surrogateEfficiency}
\vspace{1ex}
\begin{tabular}{|cc|ccccc|}\hline
\multirow{2}{*}{wire}&\multirow{2}{*}{$p$} & \multicolumn{5}{c|}{$\delta M$}\\ \cline{3-7}
                     &                    & $2$ & $4$ & $6$ & $8$ & $10$\\ \hline
\multirow{3}{*}{\num{9}}&\num{1}&\num{8}&\num{104}&\num{106}&\num{108}&\num{114}\\
                        &\num{2}&\num{13}&\num{17}&\num{21}&\num{25}&\num{29}\\ 
                        &\num{3}&\num{18}&\num{20}&\num{22}&\num{24}&\num{26}\\ \hline
\multirow{3}{*}{\num{10}}&\num{1}&\num{12}&\num{36}&\num{40}&\num{124}&\num{124} \\
                         &\num{2}&\num{13}&\num{17}&\num{21}&\num{25}&\num{29} \\
                         &\num{3}&\num{18}&\num{20}&\num{22}&\num{24}&\num{26} \\ \hline
\end{tabular}
\end{center}
\vspace{1ex}
}
For two wires, this failure probability has been computed using the iterative algorithm resulting in $P_{\F,9}^\hyb\approx\num{0.013}$ and $P_{\F,10}^\hyb\approx\num{0.007}$. 

Then, the system failure probability was estimated to ${P_{\F,\text{s}}^\hyb \leq \num{0.0200}}$.
To assess the computational savings by the usage of a surrogate model, the surrogate efficiency
has been evaluated. The main findings were that a more accurate surrogate model leads to a faster convergence of the hybrid algorithm. However, the cost of setting up higher order surrogates increases rapidly. In terms of efficiency, all investigated configurations showed significantly reduced computational cost with respect to pure Monte Carlo sampling. The drawback of the presented algorithm lies in the iteration parameter $\delta M$ that needs to be determined empirically. An alternative to the here presented iterative approach is the usage of an adjoint error approach to obtain an a posteriori estimator for the threshold $\delta$ \cite{Roemer_2017aa}. This error estimator gives full control on the accuracy of the failure probability estimate.

\medskip
\section*{Acknowledgements}
The authors would like to thank Roland Pulch for bringing the hybrid approach of Li and Xiu to their attention and for the fruitful discussions on the topic.

The work is supported by the European Union within FP7-ICT-2013 in the context of the \emph{Nano-electronic COupled Problems Solutions} (nanoCOPS) project (grant no. 619166), by the \emph{Excellence Initiative} of the German Federal and State Governments and the Graduate School of Computational Engineering at TU Darmstadt.

\end{multicols}
\end{document}